\documentclass[final,sort&compress]{aipproc}
\layoutstyle{6x9}
\usepackage{amsmath,amssymb,amsfonts}%
\usepackage{mathrsfs,bbm}%
\newcommand{\dft}{\textsl{DFT}}
\newcommand{\lda}{\textsl{LDA}}
\newcommand{\uspp}{\textsl{US-PP}}%
\newcommand{\pz}{\textsl{PZ}}
\newcommand{\pbe}{\textsl{PBE}}
\newcommand{\sr}{\textsl{SR}}
\newcommand{\fr}{\textsl{FR}}
\newcommand{\ks}{\textsl{KS}}
\newcommand{\so}{\textsl{SO}}
\newcommand{\pdos}{\textsl{PDOS}}
\newcommand{\stm}{\textsl{STM}}
\newcommand{\mcbj}{\textsl{MCBJ}}
\newcommand{\qe}{\textsc{quantum-espresso}}
\newcommand{\ang}{\ensuremath{\textnormal{\AA}}}
\newcommand{\ry}{\ensuremath{{\mathrm{Ry}}}}%
\newcommand{\ev}{\ensuremath{{\mathrm{eV}}}}%
\newcommand{\hgo}{\ensuremath{{e^2/h}}}%
\newcommand{\ef}{\ensuremath{E_{\mathrm{F}}}}%
\newcommand{\echem}{\ensuremath{E_{\mathrm{chem}}}}%
\newcommand{\dco}{\ensuremath{d_{\mathrm{C-O}}}}%
\newcommand{\dptc}{\ensuremath{d_{\mathrm{Pt-C}}}}%
\newcommand{\npt}{\ensuremath{N_{\mathrm{Pt}}}}%

\begin{document}

\title[Interaction of a CO molecule with a Pt monoatomic chain]%
{Interaction of a CO molecule with a Pt monoatomic chain: the top geometry}%


\author{G. Sclauzero}{address={International School for Advanced Studies (SISSA) and CNR-INFM Democritos National Simulation Center, Via Beirut 2-4, I-34014 Trieste, Italy}}

\author{A. Dal Corso}{address={International School for Advanced Studies (SISSA) and CNR-INFM Democritos National Simulation Center, Via Beirut 2-4, I-34014 Trieste, Italy}}

\author{A. Smogunov}{address={International Centre for Theoretical Physics (ICTP), Strada Costiera 11, I-34014 Trieste, Italy},altaddress={International School for Advanced Studies (SISSA) and CNR-INFM Democritos National Simulation Center, Via Beirut 2-4, I-34014 Trieste, Italy}}

\author{E. Tosatti}{address={International Centre for Theoretical Physics (ICTP), Strada Costiera 11, I-34014 Trieste, Italy},altaddress={International School for Advanced Studies (SISSA) and CNR-INFM Democritos National Simulation Center, Via Beirut 2-4, I-34014 Trieste, Italy}}

\begin{abstract}
  Recent experiments showed that the conductance of Pt nanocontacts and nanowires is measurably reduced by adsorption of CO. 
  We present \dft\ calculations of the electronic structure and ballistic conductance of a Pt monoatomic chain and a CO molecule adsorbed in an on-top position. 
  We find that the main electronic molecule-chain interaction occurs via the $5\sigma$ and $2\pi^{\star}$ orbitals of the molecule, involved in a donation/back-donation process similar to that of CO on transition-metal surfaces.
The ideal ballistic conductance of the monoatomic chain undergoes a moderate reduction by about $1.0\ G_0$ (from $4\ G_0$ to $3.1\ G_0$) upon adsorption of CO.
By repeating all calculations with and without spin-orbit coupling, no substantial spin-orbit induced change emerges either in the chain-molecule interaction mechanism or in the conductance.
\end{abstract}

\maketitle


\section{Introduction}

The conductance of metallic nanocontacts produced by scanning tunneling microscopy (\stm) or by mechanically controllable break-junctions (\mcbj) is routinely measured and reported in conductance histograms \citep{agrait2003}. 
The nanocontacts are much shorter in size than the electron mean free path so that the conductance is ballistic and quantized, rather than diffusive and non quantized.
Quantization is most readily observed in the form of peaks in conductance histograms obtained by superposing the conductance traces of many break junctions events. 
In some metals, such as gold and platinum, short tip-suspended monoatomic chains can be formed at the nanocontact \citep{nature1998,smit2001,rodrigues2003}; in these cases the main histogram peak with the lowest conductance is due to the atomic chain. 
This peak is usually found near $1 \;G_0$ ($G_0=2\,\hgo$) for clean Au nanowires \citep{agrait2003,nature1998} and at about $1.5 - 2.0\;G_0$ in Pt, where 5$d$ electrons contribute to the conductance \citep{untiedt2004,kiguchi:035205}. 
According to Landauer's theory, ballistic conductance is determined by electron transmission and reflection at the contact. 
The presence of small adsorbates on the nanowire or in the nanocontact may naturally change electron transmission and reflection and alter substantially the conductance. 
In the case of Pt nanocontacts this effect has been demonstrated experimentally for $\mathrm{H_2}$ \citep{smit:2002} and for CO \citep{untiedt2004,kiguchi:035205}.
For instance, in presence of CO the conductance histogram of Pt is modified to exhibit two peaks at about $0.5\ G_0$ and $1.0\ G_0$ \citep{untiedt2004,kiguchi:035205}.
Only the lower peak near $0.5\ G_0$ has been discussed theoretically \citep{strange:125424}, while that at $1\ G_0$ has not. 
At the chemical level, in the experimental work of \citet{kiguchi:035205} it was suggested that the relative strength of CO adsorption on Au, Cu, Pt and Ni nanocontacts could be explained by the \citeauthor{blyholder1964} model \citep{blyholder1964}, but the binding between a CO molecule and a Pt nanowire has not been addressed in this respect. 
Our calculations presently fill this gap. 
Finally, the effect of spin-orbit (\so) coupling on the molecule-chain binding and on the nanocontact ballistic conductance has not been taken into account in previous calculations, even though it is known that \so\ is so important to change significantly the electronic band structure of Pt monoatomic wires \citep{delin2003,dalcorso2006}.

In this work we study the interaction between CO and a Pt nanowire using a very simplified model: we consider a straight, isolated, Pt atomic chain with one molecule of CO adsorbed on-top (see the inset in Fig.~\ref{fig:cond}). 
Calculations use a periodically repeated supercell containing \npt\ atoms of Pt and the molecule with its axis perpendicular to the chain\footnote{
We verified the convergence of our results with respect to \npt\ and to the other parameters of the supercell. In particular the \pdos\ and the transmission have been calculated with $\npt=17$.
}.
We perform electronic structure calculations in the framework of density functional theory (\dft) \citep{hohenberg1964} within the local density approximation (\lda) for the exchange and correlation energy using the \pz\ functional \citep{perdew1981} (the optimized distances and the chemisorption energies have been calculated also with the gradient corrected \pbe\ functional \citep{perdew1996}).
The role of \so\ coupling is investigated by comparing the results obtained with scalar-relativistic (\sr) and fully-relativistic (\fr) ultrasoft pseudo potentials\footnote{
Within \lda, the Pt pseudopotential is that given in Ref.~\citealp{dalcorso2005}, while for C and O we generated new \uspp. With \pbe\ we generated new \sr\ \uspp\ for all the atoms (the parameters will be reported elsewhere).
We adopt a plane wave basis set with the following cut-off for the kinetic energy of the wavefunctions (charge density): \sr-\lda, $29\ (300)\ \ry$; \fr-\lda, $32\ (300)\ \ry$; \sr-\pbe\ $32\ (320)\ \ry$.}
(\uspp) \citep{vanderbilt1990}, the latter including the \so\ effect \citep{dalcorso2005}.
The ballistic conductance has been calculated with the Landauer-B\"uttiker formula, $G_0=2\,\hgo\;T(E_F)$, evaluating the transmission $T$ (obtained with the method described in Refs.~\citealp{choi1998,smogunov2004b,dalcorso2006}) at the Fermi energy \ef.
All calculations have been performed with the computer codes contained in the \qe\ package \citep{pwscf}.

\section{Electronic structure and ballistic conductance}\label{sec:main}

We optimized the platinum-carbon distance \dptc\ and the intermolecular distance \dco, while keeping the Pt atoms aligned at their theoretical equilibrium distance in the isolated chain ($2.34\;\ang$ with \lda, $2.39\;\ang$ with \pbe). 
Within \lda\ the optimal geometrical parameters are $\dptc=1.82\;\ang$ ($1.81\;\ang$) and $\dco=1.14\;\ang$ ($1.14\;\ang$) in the \sr\ (\fr) case, while the \sr-\pbe\ calculation gives slightly longer distances: $\dptc=1.84\;\ang$ and $\dco=1.15\;\ang$. 
The adsorbed C-O distance is slightly longer than the equilibrium value calculated for the isolated molecule, which is $1.13\;\ang$ with \lda\ and $1.14\;\ang$ with \pbe.
The chemisorption energy \echem\ is calculated as the difference between the energy of the optimized geometry and the sum of the energies of the isolated wire and of the isolated molecule.
With \pbe, which usually gives realistic estimates, we get $\echem=-1.4\;\ev$, while \lda, which is known to overbind, gives larger values, $\echem=-1.9\;\ev$ in the \sr\ case and $\echem=-2.0\;\ev$ in the \fr\ case. 
\begin{figure}[tb]
  \includegraphics[width=1.0\textwidth]{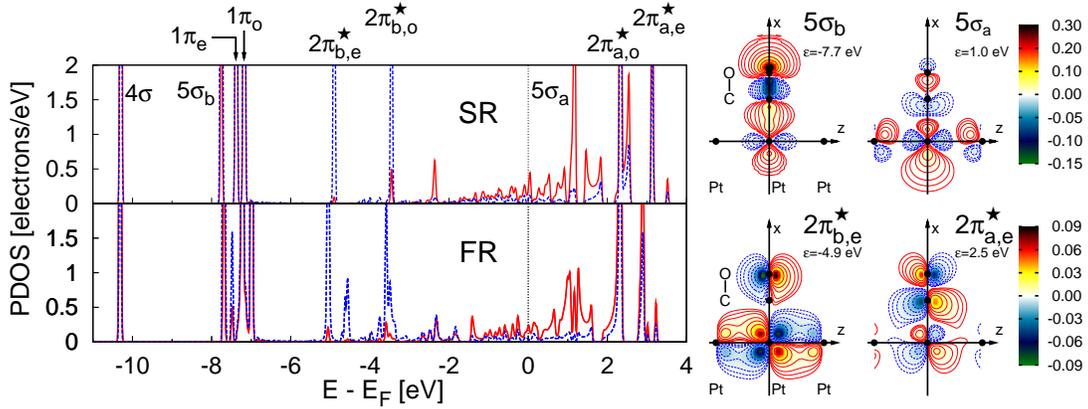}
  \caption{Left: sum of the \pdos\ on the atomic orbitals of C (solid lines) and O (dashed lines) in the \sr\ and \fr\ cases. Right: \sr-\ks\ eigenstates associated to some of the peaks in the \sr-\pdos.}
  \label{fig:pdos}
\end{figure}
Our results confirm that the interaction between the CO molecule and the Pt chain can be rationalized by the \citeauthor{blyholder1964} model (and its improvements), used to explain the chemical bond between CO and transition metal surfaces \citep{blyholder1964,fohlisch2000}. 
The CO--metal bond is characterized by a balance between donation of $5\sigma$ electrons from the molecule to the metal and back-donation from the Pt $d$-bands to the empty $2\pi^{\star}$ molecular orbital, which gets partially filled by electrons. 
This can be argued by looking at the density of states projected on valence orbitals of C and O (\pdos), shown in Fig.~\ref{fig:pdos} for the \lda\ case.
The nature of the peaks in the \sr\ \pdos\ has been identified through symmetry arguments or by directly visualizing the Kohn-Sham (\ks) orbitals. 
In addition to the sharp \pdos\ peaks due to CO valence orbitals ($3\sigma$, not shown since much lower in energy, $4\sigma$, $5\sigma$, $1\pi$ below the Fermi energy \ef\ and $2\pi^{\star}$ above \ef), we find peaks which are not present in the \pdos\ of the isolated CO and thus are due to the interaction with Pt.
In particular, the peak at $1.2\;\ev$ corresponds to a new state $5\sigma_a$ (shown in Fig.~\ref{fig:pdos}) due to antibonding hybridization between the $5\sigma$ orbital and the $s$ and $d$ bands of Pt; it lies \emph{above} \ef\ and thus gives rise to electron donation. 
The two sharp peaks at $-4.9\;\ev$ and $-3.5\;\ev$, instead, are due to bonding hybridization between the $2\pi^{\star}$ orbitals and the $d$ bands of Pt: since these new states lie \emph{below} \ef\ they are filled with metal electrons by back-donation. 
We note here that each of the $\pi$ orbitals is split into an even state ($2\pi^{\star}_{\textrm{b,e}}$ and $2\pi^{\star}_{\textrm{a,e}}$, both reported in Fig.~\ref{fig:pdos}, and $1\pi_{\textrm{e}}$) and an odd state ($1\pi_{\textrm{o}}$, $2\pi^{\star}_{\textrm{b,o}}$ and $2\pi^{\star}_{\textrm{a,o}}$) with respect to the mirror plane which contains the chain and the molecule. 
We finally point out that, as in the case of CO adsorbed on surfaces, the $5\sigma_b$ level has moved below the $1\pi$ level, and that there is the same mixing between $4\sigma$ and $5\sigma$ and between $1\pi$ and $2\pi^{\star}$ (see Ref.~\citealp{fohlisch2000}). 
Comparing the position of the peaks in the \sr\ \pdos\ with those in the \fr\ \pdos, we can argue that this model of the interaction is valid also in presence of \so\ coupling. 
In fact, although the exact position of the peaks and their intensity may vary, we can recognize in the \fr\ \pdos\ the same pattern of levels identified in the \sr\ \pdos. 
This is due to the fact that the strength of the CO--Pt interaction is much larger than the effects of \so\ coupling on this binding. 

\begin{figure}[tb]
  \includegraphics[width=0.85\textwidth]{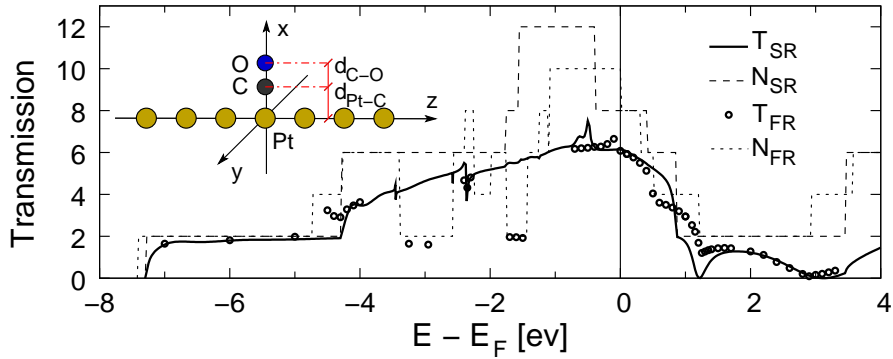}
  \caption{Electron transmission as a function of the energy. The \sr\ (\fr) trasmission $T_{\sr}$ ($T_{\fr}$) is displayed with solid lines (open circles), while the transmission of an ideal monoatomic chain $N_{\sr}$ ($N_{\fr}$), is shown with dashed (short dashed) lines. Inset: geometry and optimized distances.}
  \label{fig:cond}
\end{figure}

Fig.~\ref{fig:cond} shows the calculated transmission curve of the ideal (junction-free) Pt chain with the adsorbed CO molecule, calculated in the \sr\ and \fr\ cases within \lda.
The calculated \sr\ (\fr) conductance is about $3.1\;G_0$ ($3.0\;G_0$), reduced by about $1.0\;G_0$ relative to the isolated Pt monoatomic chain, whose theoretical conductance is $G=4\;G_0$ (see Ref.~\citealp{dalcorso2006}).
Below \ef, between $-1.6\;\ev$ and $-0.4\;\ev$ we find a larger reduction in transmission with respect to the isolated chain, because the Pt states having orbital angular momentum along the chain $m=\pm 2$ are heavily perturbed in the vicinity of CO.
Above \ef, we find a notch in the transmission curve at about $1.2\;\ev$, precisely the energy where the new $5\sigma_a$ state is formed.
The \fr\ transmission agrees quite well with the \sr\ one near \ef, since the band with total angular momentum along the chain $m_j=\pm 5/2$,  which is raised in energy by the \so\ effect, is almost completely blocked by the interaction.
In fact this band derives from the $m=\pm 2$ band, which is not well transmitted in the \sr\ case.
The largest differences between \sr\ and \fr\ transmission lie in the energy ranges $-3.4\;\ev\leqslant E \leqslant -2.6\;\ev$ and $-1.7\;\ev\leqslant E \leqslant -1.4\;\ev$ (where the different hybridization of the 
\fr\ bands depletes the number of available channels), but they are far away from the Fermi level and hence they do not produce, in this system, a measurable effect on the conductance.


\end{document}